\documentclass{llncs}

\usepackage[graphicx]{realboxes}

\usepackage{graphicx}
\usepackage{amsmath}
\usepackage{amssymb}
\usepackage{color}
\usepackage{tabularx}
\usepackage{todonotes}
\usepackage{url}
\usepackage[T1]{fontenc} 
\usepackage[utf8]{inputenc}

\usepackage{algorithm}
\RequirePackage[noend]{algpseudocode}
\usepackage{listings}
\usepackage{wrapfig}
\usepackage{stmaryrd}
\newcommand{\code}[1]{\texttt{\hyphenchar\font45\relax #1}}
\DeclareMathOperator{\ite}{ite}
\DeclareMathOperator{\bvadd}{bvadd}
\DeclareMathOperator{\add}{add}
\DeclareMathOperator{\bvmod}{bvmod}

\RequirePackage{booktabs}
\RequirePackage{longtable}
\RequirePackage{array}
\RequirePackage{tabu}
\RequirePackage{tabularx}
\RequirePackage{siunitx}
\RequirePackage{multirow}
\RequirePackage{rotating}
\title{Unbounded Software Model Checking with Incremental SAT-Solving \thanks{This work was supported by Baden-Württemberg Stiftung project HIVES}}
\author{Marko Kleine Büning, Tomas Balyo, Carsten Sinz}
\institute{
  Karlsruhe Institute of Technology (KIT), Germany\\
  \url{marko@kleinebuening.com} \\
  \url{{tomas.balyo, carsten.sinz}@kit.edu}
}

\begin{document}

\maketitle
\begin{abstract}
This paper describes a novel unbounded software model che\-ck\-ing approach to find errors in programs written in the C language based on incremental SAT-solving. Instead of using the traditional assumption based API to incremental SAT solvers we use the DimSpec format that is used in SAT based automated planning. A DimSpec
formula consists of four CNF formulas representing the initial, goal and intermediate states and the relations
between each pair of neighboring states of a transition system.
We present a new tool called LLUMC which encodes the presence of certain errors in a C program into a
DimSpec formula, which can be solved by either an incremental SAT-based DimSpec solver or the IC3 algorithm for invariant checking. 
We evaluate the approach in the context of SAT-based model checking for both the incremental SAT-solving and the IC3 algorithm. We show that our encoding expands the functionality of bounded model checkers by also covering large and infinite loops, while still maintaining a feasible time performance. Furthermore, we demonstrate that our approach offers the opportunity to generate runtime-optimizations by utilizing parallel SAT-solving. 
\end{abstract}

\section{Introduction}
\label{ch:Introduction}
Software has become an important part of almost all modern technical devices, such as cars, airplanes, household appliances, therapy machines, and many more. The cars of tomorrow will drive on their own but will be controlled by software. As shown by serious accidents like the rocket crash of Ariane flight 501 \cite{lions1996ariane}, the massive overdoses of radiation generated by the therapy machine Therac-25 \cite{leveson1993investigation} or the car crash of the Toyota Camry in 2005 \cite{koopman2014case} software is never perfect, it almost always contains errors and bugs. 
While testing of software can only cover a limited number of program executions, software verification can guarantee a much higher coverage while producing proofs for the existence or absence of errors. There exist several different software verification approaches, as for instance symbolic execution \cite{khurshid2003generalized} and bounded model checking \cite{biere2003bounded}. Boun\-ded model checking inlines function calls and unrolls loops a finite number of times, say $k$-times, where $k$ is called the bound of the program. This unrolling reduces the complexity of the problem to a feasible level, though it limits the coverage and precision of these approaches.

By means of extending the functionality of bounded model checkers, we developed a novel unbounded model checking approach. To this end, we removed the bound that limits all bounded model checkers and created a transition system that is traversed by an incremental SAT-solver or an invariant checking algorithm. 
We focus on sequential programs written in C and use the low-level code representation of the compiler framework \emph{LLVM} as an intermediate language. Based on this representation we derived an encoding of the program verification task into a DimSpec formula. A DimSpec formula uses four CNF formulas to specify a transition system and is often used in SAT based automated planning. We first encode the program into an SMT formula and, subsequently, we generate the SAT-problem in DimSpec format. The resulting DimSpec formula is then solved by either an incremental SAT-solver that unrolls the transition system to find a transition path to the error state or an invariant checking algorithm that refines an over-approximation of a transition path to the error state. 

Our verification system uses Clang and LLVM version 3.7.1 to compile C-code into the LLVM intermediate language. Then our new tool LLUMC (Low-Level-Unbounded-Model-Checker) generates DimSpec formulas representing the presence of certain errors in the program. To solve the generated formulas \cite{gocht2017Incre} we either use the incremental SAT-solver IncPlan \cite{balyo2017IncPlan} or the invariant checking algorithm implemented in the solver MinireachIC3 \cite{balyo2016reachlunch}. 
LLUMC was inspired by the bounded model checker LLBMC  \cite{merz2012llbmc} but runs independently.
Our evaluation is based on the Software Verification Competition (SV-Comp) and shows the correctness and feasibility of our approach. LLUMC is available online at \cite{llumc2017}. 

\section{Preliminaries}
\label{ch:Preliminaries}
\label{sec:Preliminiaries:Boolean Satisfiability Problem (SAT)}
We assume the reader is familiar with propositional logic, first-order-logic and SAT and use definitions and notations standard in SAT. This section will introduce incremental SAT-solving and describe the theory of bit-vectors in the context of SMT-solving. Furthermore, the software bounded model checking approach is briefly described. 

\subsubsection{Incremental SAT-Solving}
In the \emph{assumption based interface} \cite{een2003extensible}, two methods are used:
\code{add}(C)\, and \code{solve}(A), where $C$ is a clause and $A$ a set of literals called assumptions.
All clauses can be added with the \code{add} method and their conjunction can then be solved under the condition that all literals in $A$ are true by $\code{solve}(A)$. 
To add a removable clause $C$, we add $(C \lor a)$, where $a$ is an unused variable. The clause is only relevant, if we add the literal $\lnot a$ (called activation literal) to the assumptions $A$. If the activation literal is not added to the assumptions $C$ is essentially removed from the set clauses.

\noindent\textbf{DimSpec Formulas}
A DimSpec formula represents a transition system with states $t_0, t_1, \ldots, t_k$, where each state is a full truth assignment on $n$ Boolean variables $x_1,\ldots,x_n$. It consists of four CNF formulas: $\mathcal{I}, \mathcal{U}, \mathcal{G}$ and $\mathcal{T}$, where $\mathcal{I}$ are the initial clauses, i.e., clauses satisfied by $t_0$, $\mathcal{G}$ are goal clauses satisfied by final state $t_k$, the $\mathcal{U}$ clauses are satisfied by each individual state $t_i$ and finally the transitional clauses $\mathcal{T}$ are satisfied by each pair of consecutive states $t_i, t_{i+1}$. The clause sets $\mathcal{I}, \mathcal{U}, \mathcal{G}$ contain the variables $x_1,\ldots,x_n$ and $\mathcal{T}$ contains $x_1,\ldots,x_{2n}$. Testing whether the goal state is reachable from the initial state within $k$ steps is equivalent to checking whether the following formula $F_k$ is satisfiable.
\begin{equation*}
F_k = \mathcal{I}(0) \land \left(\bigwedge_{i=0}^{k-1}\bigl(\mathcal{U}(i)\land \mathcal{T}(i,i+1)\bigr)\right)\land \mathcal{U}(k) \land \mathcal{G}(k)\, ,
\end{equation*}
where $\mathcal{I}(i)$, $\mathcal{G}(i)$, $\mathcal{U}(i)$ and $\mathcal{T}(i, i+1)$ denote the respective formulas where each variable $x_j$ is replaced by $x_{j+i*n}$. One way to find the smallest number of steps to reach the goal state from initial state is to solve $F_1, F_2, \ldots$ until a satisfiable formula is reached.
An efficient way to implement this is to use an incremental SAT solver with the assumption based interface via the following steps.
\begin{alignat*}{2}
\mathrm{step}(0):\quad &\code{add}(\mathcal{I}(0) \land (a_0 \lor \mathcal{G}(0))\land \mathcal{U}(0))\\
&\code{solve}(assumptions=\{\lnot a_0\})\\
\mathrm{step}(k):\quad &\code{add}(\mathcal{T}(k-1,k) \land (a_k \lor \mathcal{G}(k)) \land \mathcal{U}(k) ) \\
&\code{solve}(assumptions=\{\lnot a_k\})\, .
\end{alignat*}
This algorithm works only if the goal state is reachable from the initial state, otherwise it does not terminate. A more sophisticated approach that can detect unreachability is described next.

\noindent\textbf{IC3 algorithm}
A different approach to solve a transition system reachability is described in \cite{bradley2011sat} and implemented in the tool IC3 (\textbf{I}ncremental \textbf{C}onstruction of \textbf{I}nductive \textbf{C}lauses for \textbf{I}ndubitable \textbf{C}orrectness). Given a transition system $S$ and a safety property $P$ the algorithm can prove that $P$ is $S-invariant$, meaning that regarding $S$ the property $P$ is true in all reachable states or produces a counterexample. IC3 incrementally refines a sequence of formulas $F_0, F_1, ..., F_k$ that are over-approximations of the set of states reachable in at most $k$ steps. It can extend the formula sequence in major steps that increase $k$ by one. In minor steps the algorithm refines the approximations $F_i$ with $0 \leq i \leq k$ by conjoining clauses to $F_0, \ldots, F_j$ with $0 \leq j \leq k$. 
Given a finite transition system $S$ and a safety property $P$, the IC3 algorithm terminates and returns {true}, iff $P$ is true in all reachable states of $S$ \cite{bradley2011sat}. The IC3 algorithm was implemented and adjusted\footnote{The clause sets $\mathcal{I}, \mathcal{U}, \mathcal{T}$ represent the transition system $S$ and $\mathcal{G}$ represents the negation of the invariant property $P$.} 
to the DimSpec format in the tool MinireachIC3 by Suda \cite{balyo2016reachlunch}.

\subsubsection{Satisfiability Modulo Theories (SMT)}
\label{sec:Preliminiaries:Satisfiability Modulo Theories (SMT)}
Due to quantifiers, first-order-logic is generally undecidable but there are numerous decidable subsets. The problem of solving those subsets or theories is called satisfiability modulo theories or SMT. There is a lot of research on various theories, there are for example the theory of arrays, bit-vectors, floating points, heaps, linear arithmetic and many more. These theories can be seen as restrictions on possible models of first-order-logic formulas \cite{merz2016theory}. In this paper, we will restrict ourselves to the theory of bit-vectors.
SMT was standardized by the SMT-LIB initiative \cite{barrett2010smt}. We will use the same notations, especially when referring to SMT functions defined in the different theories. Such an SMT-LIB function could for example be $\bvadd(b_1,b_2)$, describing the addition of two bit-vectors $b_1$ and $b_2$. A more complex function is called \code{if-then-else} ($\ite$) and is defined by:
\begin{equation}
\label{eq:ite}
\forall c,x,y,z \left(x=\ite(c,y,z)\Leftrightarrow c \land x = y \lor \lnot c \land x = z\right)\, .
\end{equation}

\label{sec:Preliminiaries:Theory of Bitvectors}
We refer to the \textbf{theory of fixed-size bit-vectors} defined by the SMT-LIB standard in \cite{barrett2010smt}. The theory of bit-vectors models finite bit-vectors of length $n$ and operations on these vectors into first-order-logic. The set of function symbols contain standard operations on bit-vectors as for example the addition,  multiplication, unsigned division, bit-wise and, bit-wise or, bit-wise exclusive or, left shift, right shift, concatenation, and extraction of bit-vectors. 

\subsubsection{Software Bounded Model Checking}
\label{sec:Preliminiaries:Software Bounded Model Checking}
\label{sec:Preliminiaries:LLBMC}
The general idea of bounded model checking (BMC) is to encode the states of a system and the transition between them. Furthermore, you unroll any loop and function calls $k$-times. The number $k$ is called the \emph{bound} and is the reason for the decidability of bounded model checking but also for its limitations. 
After the unrolling and encoding of the program, a formula that represents the negation of a desired property is added and the formula is solved with a SMT or SAT-solver. If the solver finds a model for the formula, the approach has found an error and the model can be used as a counterexample. The loop-bound can be increased step by step until a fixed bound $k$ is reached. Thus, the counterexample is always minimal and easier to comprehend for a user. The question to which bound the loop should be unrolled is complex and further discussed for example by Biere et al. \cite{biere2003bounded}.

Bounded model checking is implemented for example in the tool LLBMC (Low-Level-Bounded-Model-Checker). It was developed at the research group "Verification meets Algorithm Engineering" at the KIT with the aim to verify safety-critical embedded systems \cite{merz2016theory}. To support large parts of the C and C++ languages it uses the compiler framework \emph{LLVM} as it's foundation. 
With it's algorithm LLBMC is able to create very positive results and earned a number of gold, silver and bronze medals in the Software Verification Competition (SV-Comp), which we will describe and refer to in our evaluation in Section \ref{ch:Evaluation}. We will use LLBMC as a state-of-the-art reference to compare it to our approach. 

\subsubsection{LLVM Representation}
\label{sec:LLVM representation}
LLVM is an open source compiler framework project that consist of a "collection of modular and reusable compiler and tool-chain technologies" \cite{llvm}. It supports compilation for a wide range of languages and is known for its research friendliness and good documentation. To work directly on C-code is very complex and it is nearly impossible  to support all features and libraries. Thus, we use the intermediate language of LLVM, which describes the statements more directly and provides a number of optimizations and simplifications.  
\label{sec:LLVM representation:Structure}
We define a LLVM-module bottom up. The smallest executable unit is called an \emph{instruction}. An instruction is an atomic unit of execution that performs a single operation. 
A basic block is a linear sequence of program instructions having one entry point and one exit point. It may have many predecessors and many successors and may be its own successor.
The last instruction of every basic block is called \emph{terminator}. Every basic block is part of a \textit{function}. A function $(n,B,e)$ is a tuple of a name $n$, a sequence of basic blocks $B=(b_0, b_1, ..., b_m)$, and an entry block $e\in B$.
Hereinafter, we will denote the main function of every program with $f_{main}$.
A module $m=(F_m,G_m)$ is a pair of a set of function symbols $F_m$ and a set of global variable symbols $G_m$. 

\label{sec:LLVM representation:Passes and Optimizations}
To optimize our encoding, we run some predefined optimization passes from LLVM and LLBMC on the generated LLVM-module. Among other things, these optimizations remove undefined behavior in C-code, promote memory references to register references and inline the program into one main function. These optimizations are described in more detail in \cite{kleinebuening2017Unbound}. 
The resulting LLVM-module is then used as input for our encoding. 

\section{LLUMC Encoding}
\label{sec:LLUMC encoding}
\label{sec:Overview and Theory}
A bug or error in a software program is a well-known notion but there exists no universal definition. A general concept is that a program has an error, if it does not act according to its specification. For our approach this definition is not specific enough. We will not cover all possible errors but concentrate on two main properties. 
One of them is the occurrence of an \emph{undefined overflow} for the signed arithmetic operations addition, subtraction, multiplication and division. We define undefined overflows independent from the variable type and thus independent of the bit-vector length representing the variable. Let $v$ be a variable in two's complement and let $\ell$ be the bit-length of $v$, then $max_v$ returns the maximal value for $v$: $2^{\ell-1}-1$ and $min_v$ returns the minimal value $-2^{\ell-1}$. 
In the C language unsigned overflows are defined by a wrap around. The addition of two unsigned integers $\langle x\rangle ^u_I$ and  $\langle y\rangle ^u_I$ is  e.g. defined modulo $max\_int$:
\begin{equation*}
\langle x\rangle ^u_I +  \langle y\rangle ^u_I = x + y\mod max\_int+1\,.
\end{equation*}
Thus, we can consider undefined overflows solely on signed variables.  
\begin{definition}[Undefined Overflow]
	\label{def:undefined Overflow}
	Let $\langle x\rangle^s_l, \langle y\rangle^s_l$ be signed variables of length $\ell$, then an undefined overflow occurs, if 
	\begin{enumerate}
		\item $\langle x\rangle^s_\ell+\langle y\rangle^s_\ell > max_\ell$,
		\item $\langle x\rangle^s_\ell-\langle y\rangle^s_\ell < min_\ell$,
		\item $\langle x\rangle^s_\ell \cdot \langle y\rangle^s_\ell > max_\ell \text{ or } \langle x\rangle^s_\ell \cdot \langle y\rangle^s_\ell < min_\ell$,
		\item $\langle x\rangle^s_\ell \div \langle y\rangle^s_\ell$ with $\langle x\rangle^s_\ell=min_\ell$ and $\langle y\rangle^s_\ell=-1$.\\
	\end{enumerate} 
\end{definition}
The other property for our error definition, regards calls to \emph{assume} and \emph{assert}. A program acts according to its specification, if the assert statements are true under the condition that the assume conditions are met. If the assume condition is not met, the further run of the program is not specified and thus no errors can occur. 
With these two properties in mind, we can define the term error for our approach. 
\begin{definition}[Program Error in LLUMC]
	\label{def:Error}
	Let $p$ be a program. Then there exists an error in $p$, if all calls to assume that are prior to an assert statement or possible overflow are true and one of the following holds. 
	\begin{enumerate}
		\item An assertion is false: a call to assert with the parameter false. 
		\item The occurrence of an undefined overflow for an arithmetic operation.
	\end{enumerate}
\end{definition}
Of course, there are other errors that can happen during a program execution like irregular bit-shifting, non-termination and many more. These errors can be regarded in future work and for the remainder of this paper the expression "error" is equated with the above definition. 

To find these errors we regard an LLVM-module as stated in Section \ref{sec:LLVM representation:Passes and Optimizations}. After inlining all function calls, we can concentrate on just the main function. Every basic block together with its variable assignment can be seen as a \emph{state}. We then add a special \emph{error state} and try to find a path from the entry state, defined by the entry block of the main function, to the error state. Therefore, we first define the state space of our encoding. 

\subsubsection{State Space}
\label{sec:LLUMC encoding:State Space}
Transition from one state to the next state will always represent the transition from one basic block to the next with respect to its current variable assignment. Often this kind of encoding is called \emph{small block encoding} \cite{beyer2009software}. 
According to the theory of bit-vectors, we define every state variable as a bit-vector of length $n$. The number of bit-vectors in the state, including the bit-vectors representing the current and previous basic block, define the number of SMT variables that are needed to encode the state and the number of bits in total represent the number of CNF variables needed.

The focus on the theory of bit-vectors, allows us to ignore the state of the main memory and concentrate on the immediate LLVM-module\footnote{Generally, encoding the state of the main memory is not easily realized and to integrate a main memory model in our approach requires further research.}. First of all, every state has to save the current basic block. Hereinafter $|B|$ denotes the number of basic blocks of the main function. For our encoding we need two additional blocks. The \emph{ok} block represents a safe state from which on, no more errors can occur. This block is reached when the program terminates with the output 0 or when an assume condition is not met. The second block is called \emph{error} and is our goal state, representing that an error occurred. With the function $enc(bb):Basic Block \rightarrow \mathbb{N}$ we uniquely map every basic block to a natural number. If there are $|B|$ basic blocks in $main$, then the bit-vector needs to have the length $\lceil \log_2(|B|+2)\rceil$ to encode the current basic block. We call this variable: 
\begin{equation*}
curr=\left(curr_1, curr_2, ..., curr_{\lceil \log_2(|B|+2)\rceil}\right) \text{, for Boolean variables }curr_i.
\end{equation*}
In LLVM the value of a register can depend on the previous basic block and must thus also be encoded:
\begin{equation*}
pred=\left(pred_1,pred_2, ..., pred_{\lceil \log_2(|B|+2)\rceil}\right) \text{, for Boolean variables }pred_i.
\end{equation*}
Furthermore, we need to save the current variable assignment. We do not need the assignment of all variables, but should concentrate on those that will be accessed later on and cannot be optimized away. Those variables can be classified by two properties and we call the set of those variables $V$:
\begin{enumerate}
	\item Variables that are used in more than one basic block and 
	\item variables that are read before written in the same basic block, which is part of a loop.
\end{enumerate}
It is enough to add only those variables to the state space, because all other variables are included during the encoding of the entailing basic block and their value is not directly used for a transition step. The length of the variables depends on their type. The standard integer in C has a width of 32 bits, long has 64 and Boolean values have a width of 1. There are other types but their lengths is always specified by LLVM and can thus be easily extracted. 

\begin{definition}[State]
	\label{def:state}
	The state space is the set of bit-vector variables: $State\-Space = \{curr,\ pred\} \cup V\,.$ Every variable of the state space has a fixed bit-length $\ell$ and can take on concrete bit-vectors of length $\ell$ as values. For a specific time point $k$ the state \emph{state(k)} is the assignment of concrete bit-vectors to every variable. 
\end{definition}
\subsubsection{Encoding to SMT}
\label{sec:LLUMC encoding:Encoding to SMT}
Our aim is to develop an encoding for an LLVM-module defined in Section \ref{sec:LLVM representation:Passes and Optimizations} that fits the DimSpec format. Therefore, we must define the four CNF formulas $\{\mathcal{I},\mathcal{G}, \mathcal{U}, \mathcal{T}\}$ in such a way that if there exists a transition from $\mathcal{I}$ to $\mathcal{G}$ defined by $\mathcal{T}$ and restricted by $\mathcal{U}$ then there exists an error in the given program code. 

The initial formula $\mathcal{I}$ can be created by encoding the entry block of the LLVM-module. Due to the restriction on the theory of bit-vectors global variables are not regarded, because they always include a memory access. The encoding has to represent the state that we are currently at the first basic block and that there were no prior actions. We declare the entry block itself as the predecessor to exclude any prior actions. The initial formula is thereby time-independent, because the entry block is the same for every time step. The rest of the variable assignment is arbitrary at this point and can be left undeclared.   
\begin{definition}[Encoding of Initial Formula]
	Let \emph{entry} be the name of the first block, then the initial formula $\mathcal{I}(k)$ for the LLVM-module and for $k\in\mathbb{N}$ is defined as: 
	\begin{alignat*}{2}
	curr 		&= \emph{enc}(entry)\quad \land  \\
	pred	&= \emph{enc}(entry)\,.
	\end{alignat*}
\end{definition}
The encoding of the goal formula $\mathcal{G}$ is also time-independent and can be defined accordingly. 
\begin{definition}[Encoding of Goal Formula]
	Let \emph{error} be the name of the error block, then the goal formula $\mathcal{G}(k)$ for the LLVM-module and for $k\in\mathbb{N}$ is defined as: 
	\begin{equation*}
	curr = \emph{enc}(error)\,.
	\end{equation*}
\end{definition}
The universal formula consists of constraints that have to be true in all states. In our case, that are boundaries for the variables $curr$ and $pred$. In the previous section, the number of bits needed to encode the current and previous basic block were defined as $|B|+2$. In most cases $|B|+2$ is not a power of two and thus bigger numbers can be represented. These numbers must be excluded at all times in the universal formula $\mathcal{U}$. 
\begin{definition}[Encoding of Universal Formula]
	Let $|B|$ be the number of basic blocks in the LLVM-module, then the universal formula $\mathcal{U}(k)$ for $k\in\mathbb{N}$ is defined as: 
	\begin{alignat*}{2}
	curr	&\leq \left(|B|+2\right) \quad \land \\
	pred	&\leq \left(|B|+2\right)\,.
	\end{alignat*}
\end{definition}

At last, we have to define the transition formula. It represents the transition between state $k$ and state $k+1$. It is important to notice that the transition formula has twice as much variables as the other formulas. To distinguish between the variables in time-point $k$ and $k+1$ every variable $v$ of our state space is called $v'$ at time-point $k+1$. Otherwise, every transition formula would be evaluated to false and thus no transition step could ever be taken. In general, the encoding of one transition has the form: 
\begin{equation}
\label{eq:transition}
state(k)\Rightarrow state(k+1)\,.
\end{equation}
We call $state(k)$ antecedent and $state(k+1)$ consequent. For each $state(k)$ that is reachable from our initial state, a transition must be defined. An undefined transition leads to an undefined $state(k+1)$ with arbitrary values. Thus, if there is a reachable, undefined transition all goal states can be reached. For the same reason, we determine that for each $state(k)$ the transition must be explicit. 
Variables that are not important for the transition should not be declared in the antecedent but should be specified in the consequent to avoid undefined values. We will use the auxiliary function
\begin{equation*}
same(bb): \text{Basic Block} \rightarrow \text{SMT-formula}
\end{equation*}
to encode that variables which are not modified in a basic block maintain their current value. The function $same(bb)$ returns the conjunction of all $var = var'$,  for all variables in our state space, that have not been modified in the transition of our basic block $bb$.\smallskip

To encode the transition between steps, we take a closer look at the current basic block, further denoted as $bb$ and customize Equation \ref{eq:transition} for different branching possibilities. We divide basic blocks into three groups and distinguish them by means of their terminator. Afterwards, we will have a special look at the function calls of $assume$, $assert$ and $exit$. These function calls together with the possibility of overflows will extend the encoding. The three different types of terminator instructions are called unconditional branching, conditional branching and return.  \smallskip

\noindent\textbf{Unconditional branching ($\textbf{br\ \%bb2}$):}
Branches to the basic block with the label $bb2$ and creates a transition from the current basic block to $bb2$.  
If the current basic block has no other instructions, only the change of basic block and the saving of the predecessor have to be encoded. Furthermore, we have to state that no variables have changed during this transition:
\begin{equation}
\label{eq-unconditional branching}
curr=\emph{enc}(bb) \Rightarrow curr'=\emph{enc}(bb2) \land pred'=\emph{enc}(bb) \land same(\emptyset)\,.
\end{equation}
This encoding is rarely complete, because it does not regard all other instructions in the basic block $bb$. Let $rl_{bb}$ be the ordered list of instructions from bottom to top in $bb$. Then we iterate over $rl_{bb}$ and regard all instructions that (1) are part of our state, (2) have not been visited before and (3) are not the terminator instruction. 
The instruction is encoded according to its type and its operands. When an instruction like $\%tmp3\, =\, \add\, i32\ 10,\, \%tmp2$ is encoded, the algorithm checks the operands first. When regarding the value $\%tmp2$, the algorithm checks whether it is a variable that is part of our state or a value calculated by an instruction, which the algorithm has to encode recursively. The stop criterion is always the occurrence of a state variable, a constant like for example $10$ or a call to assert, assume or error. For the add instruction the encoding would result in $tmp3' = \bvadd\left(10, tmp2\right)$. This generated SMT formula is then conjuncted with the consequent of equation \ref{eq-unconditional branching}. For arithmetic operations an additional overflow check formula, which is described later on, is inserted. The algorithm continues by iterating further through the list $rl_{bb}$ until there are no instructions left. \\
\noindent\textbf{Conditional branching ($\textbf{br\ \%cond,\ \%bb1,\ \%bb2}$):}
Creates a transition to $bb1$ with the condition $cond=1$ and a transition to $bb2$ with the condition $cond=0$. 
Every conditional branch has a branching condition represented as a variable ($cond$). We can extract that condition by visiting and encoding the variable representing the branching condition. 
In LLVM the branching condition is a Boolean value that is assigned by the so called $icmp$\,-$instruction$. This instruction returns a Boolean value based on the comparison of two values and it supports equality, unsigned and signed comparison. The icmp-instruction is then encoded recursively by visiting its two operands with the same approach as described for the unconditional branching. The result could for example be the condition $tmp2 > 10$.
Based on this condition, the algorithm creates two separate transitions.
\begin{alignat*}{3}
&curr = \emph{enc}(bb) \land  visitInst(cond) \Rightarrow \\
&\qquad\qquad\qquad\qquad curr'=\emph{enc}(bb1) \land pred'=\emph{enc}(bb) \land same(\emptyset)\,.\\
&curr = \emph{enc}(bb) \land \lnot(visitInst(cond)) \Rightarrow\\
&\qquad\qquad\qquad\qquad curr'=\emph{enc}(bb2) \land pred'=\emph{enc}(bb) \land same(\emptyset)\,.	
\end{alignat*}
\noindent\textbf{Return value (\textbf{$ret\ val$}):}
The value $val$ can be an arbitrary integer and represents the return value of the program as usual. This terminator creates a transition to $ok$. In an extended and already implemented version, another check is inserted verifying that the result value of a correct program is 0 and if this does not hold a transition to $error$ is created.\smallskip

Now we have to look at the calls to assume, assert, error and the possibility of overflows. During the instruction iteration of a basic block, we regard these instructions differently because they lead to a split of our transitions.\smallskip

\noindent\textbf{Method calls (\textbf{error, assume, assert}):} 
If the $error$-method, which is used to specify program errors in C-code, is called inside a basic block, we do not have to regard any other instructions and thus delete all other transitions from this basic block. We produce a single transition: 
\begin{equation*}
\label{eq:visitError}
curr=\emph{enc}(bb) \Rightarrow curr'=\emph{enc}(error) \land same(\emptyset)\,.
\end{equation*} 
The other three possibilities lead to a split of our transitions similar to the conditional branching. A call of $assume(var)$ divides the set of current transitions for our basic block. The condition is $var=0$ and leads to a transition to the $ok$ state with  $s'=\emph{enc}(ok)$. The call to $assert(var)$ is similar only with the transition to $s'=\emph{enc}(error)$ if $var=0$ holds true. In both cases, the encoding continues normally with the next instruction if the conditions are not met.\smallskip

\noindent\textbf{Overflow Checks:} 
While calls to error, assume and assert are explicit calls in the LLVM-module, we have to recognize possible overflows while still encoding the operations correctly. Therefore, an overflow check is always inserted when $visitInst(I)$ is called on an arithmetic operation with the flag $nsw$. In this case, we know that there is a signed operation with no defined wrap around. If the condition $cond_{ov}$ for an overflow is true, we transition to the $error$ state. We will give the formula for the signed addition, the formulas for subtraction,  multiplication and division are similar and comply with the undefined overflow in Definition \ref{def:undefined Overflow}.

\textbf{Addition}: The result of adding up two positive numbers must always be positive and the addition of two negative values must always result in a negative value. Whether the result is positive or negative can be seen by the sign-bit. Starting with 0, we will refer to a single bit at position $i$ of a bit-vector $b$ by $b[i]$. The position of the sign-bit has the special index $sb$. Let $res$ be the result of adding the two bit-vectors $a_1$ and $a_2$, then the condition $cond_{ov}$ for an undefined overflow is defined by:
\begin{alignat*}{2}
1=\left(\big(\lnot res[{sb}] \land a_1[{sb}] \land a_2[{sb}]) \lor (res[{sb}] \land \lnot(a_1[{sb}] \land a_2[{sb}])\big)\right)\,.
\end{alignat*}
All components of the transition formula have now been discussed. To obtain the complete transition formula the algorithm has to iterate over all basic blocks of the main function. Depending on their terminator instruction, every basic block has do be encoded according to the definitions above. To predict which transition is taken in which step would be equal to solving the whole formula. Thus, the transition formula is time independent and the transition possibilities for all time steps are part of the formula. 

\begin{definition}[Encoding of the Transition Formula]
	Let $BB$ be the set of all basic blocks of $f_{main}$ and let $encode(b)$ with $b\in BB$ be the encoding as shown above, then the transition formula $\mathcal{T}(k, k+1)$ for $k\in \mathbb{N}$ is defined by:
	\begin{equation}
	\bigwedge_{b \in BB} encode(b)\,.
	\end{equation}
\end{definition}

\begin{claim}
	\label{th:errorPathCorr}
	There exist an error as defined in Definition \ref{def:Error} for the program $p$, iff
	\begin{enumerate}
		\item $p$ is transformed into a LLVM-module $\ell$ as described in Section \ref{sec:LLVM representation} and
		\item there exists a transition path in $\ell$ from the initial state to the goal state while the universal formula holds in all states. 
	\end{enumerate}
\end{claim}
\textbf{Proof idea}: 
We forego on a formal proof, because it would require a structural induction over huge sets of C-Code and the LLVM-language.  Instead, we present short arguments and references for our claim. \\
(1): Using LLVM as a representation for C-code is widely accepted and used in research and industry. We assume that the transformation from C-code into a LLVM-module does not remove or add any errors based on the high number of research papers \cite{albarghouthi2012ufo,babic2007structural,beyer2013second} and tools like LLBMC \cite{merz2016theory} and SeaHorn \cite{gurfinkel2015seahorn}. \\
(2): The error node has three types of incoming edges: from an assert statement, from an overflow check and an edge from the error node itself. We disregard the edge that points to itself and are left with the two options that match the properties defined in \ref{def:Error}. If the encoding of the variables is, as we claim, correct and our state space is closed under $\mathcal{T}$ and $\mathcal{U}$ we can assume that the a transition path from the initial state to the error state complies with an error in the LLVM-module. 

\subsubsection{From SMT to SAT formula}
\label{sec:LLUMC encoding:From SMT to SAT formula}
The encoding of the LLVM-module gives us four SMT formulas. These formulas have to be translated into CNFs. The most widespread approach to transform SMT to CNF formulas is called \emph{bit-blasting}. We have taken one approach implemented in STP \cite{ganesh2007decision} and the ABC-library \cite{jha2009beaver} and modified these algorithms to correspond to some technical requirements of the DimSpec format. Finally, a CNF in the DimSpec format is created that can be used as input for a number of SAT-solvers.  

\section{Experimental Results}
\label{ch:Evaluation}
\label{sec:Eval:Impl}
The LLUMC-approach is implemented as a tool-chain.
First, the input file in C-code is compiled with Clang (version 3.7.1) and then optimized with LLVM and LLBMC passes. This optimized LLVM-module serves as input for the program LLUMC, which performs the encoding as described above. To transform the created SMT formulas into CNF formulas in DimSpec format, the tool STP was modified. The final renaming and aggregation is implemented directly in LLUMC. Thus, the tool produces a single CNF file in DimSpec format.

We tested two different approaches to solve the generated DimSpec/CNF formulas. The tool IncPlan \cite{gocht2017Incre} was developed at KIT and implements the incremental SAT-solving described in Section \ref{sec:LLVM representation}. It can be used with every SAT-solver that accepts the Re-entrant Incremental Satisfiability Application Program Interface (IPASIR). We have tested IncPlan with a number of SAT-solvers including Minisat \cite{sorensson2005minisat}, abcdSat \cite{chen2015minisat}, Glucose \cite{audemard2014glucose} and Picosat \cite{biere2008picosat}. While Glucose and Minisat produced good results for some benchmarks, they crashed for a number of benchmarks and thus we concentrated on the usage of abcdSat and PicoSat. 
The IC3 algorithm was implemented and adjusted to the DimSpec format in the tool MinireachIC3 by Balyo and Suda \cite{balyo2016reachlunch}. The safety property $P$ expresses that the error state should not be reachable and thus $P$ is given by $\lnot G$. Thus, we are not only able to prove the existence of errors but also their nonexistence. 

\subsection{Benchmarks}
\label{sec:Eval:Bench}
We evaluated our approach using benchmarks from the Software Verification Competition (SV-COMP) \cite{beyer2013second}. The SV-Comp is an annual competition for academic software verification tools, with the aim to compare software verifiers. The competition is conduced every year since 2012. The verification tasks are divided in different topics and verification tasks are contributed by a number of research and development groups. While we were not able to participate in the competition, the collected benchmarks serve as an excellent evaluation basis for every verifier. All benchmarks are available at \cite{svBenchmarks17} and we regarded the sub-folder \emph{c} with programs written in the language C.

We screened these benchmarks for tasks that match our theory of bit-vectors. We excluded all benchmarks that do not match our theory and removed benchmarks that include memory accesses or floating point arithmetic. Furthermore, we checked that all instructions used in the examples were implemented in LLUMC. It is notable, that nearly all instructions were implemented and only the truncate instruction, which cuts the length of values, restricts the usable benchmarks. The truncate instruction is not included in most theories of bit-vectors e.g in tools like LLBMC, because on a programming level there is not enough (signedness) information about the bit-vector to truncate it easily. Lastly, we excluded recursive and concurrent tasks due to the inlining in our approach. 

We evaluated our approach on 14 incorrect and 10 correct programs. Our approach creates a CNF formula representing the problem of finding a transition path to the error state. Thus, the desired result of our approach should be \emph{sat} in case there exists an error and \emph{unsat} if there is none. Whereby most benchmarks are smaller and have the purpose of demonstrating the correctness of our approach, we were also able to evaluate our approach for some larger problems. The benchmarks vary between 14 and 646 lines of code (LoC) and 151 to 116777 number of clauses.
The evaluation was performed on a system with 64 CPUs with 2.4GHz from which, for our sequential approach, only one was used and 483 GB memory. Each benchmark had a time limit of 600 seconds and a memory limit of 8 GB. The time needed to generate the CNF formula and to read and write CNF formulas in and out of files is negligible for larger problems. Thus, we decided to measure only the CPU time needed to solve the generated CNF formulas. 

Table \ref{tab:IncPlan} displays the result of solving the generated DimSpec/CNF formulas both with the tool IncPlan and MinireachIC3. The results of running IncPlan with the SAT-solver abcdSat were most stable and are thus displayed. One can see, that our approach generates correct encodings of the C-code and that IncPlan is able to find a satisfying model representing a transition path to the error state for erroneous programs. We also recognize that for small problems the time and memory needed is insignificant and for larger problems it is still manageable.
For programs without an error we are not able to prove anything, but the timeouts indicate the correctness of our encoding. The \emph{jain} benchmarks show that the number of iterations the SAT-solvers are able to perform in the given time depends on the complexity of the individual basic block and varies for all benchmarks.

\begin{table}
	\caption{Runtime data for Benchmarks from the SV-Comp run with IncPlan and MinireachIC3, where the column \textbf{Its} stands for the number of incremental solving steps (Iterations) performed by the SAT-solver, \textbf{ans} is the result (TO meaning timeout) and \textbf{T} and \textbf{M} represent the runtime in seconds and memory consumption in megabytes.}
	\label{tab:IncPlan}
	\centering
	\resizebox{\textwidth}{!}{
		\begin{tabular}{| l l|c r r r||c r r r |}
			\multicolumn{2}{c}{} & \multicolumn{4}{c||}{{abcdSAT}} & \multicolumn{4}{c}{{MinireachIC3}}\\
			\cline{3-10}
			\multicolumn{1}{c}{} &	\textbf{Benchmark} 	& \textbf{ans} 	 	& \textbf{T} & \textbf{M}& \textbf{Its} & \textbf{ans} 	& \textbf{T} 	& 	\textbf{M} 	& \textbf{Its}\\ \cline{2-10} \cline{1-10}
			\multirow{13}{*}{\rotatebox[origin=c]{90}{error}}
			&diamond false unreach call2 					& sat 			& 3.95& 	130		& 27				& TO	& 600			& 	241.2 		& /	\\\cline{2-10}	
			&implicitunsignedconversion false unreach call	& sat 			& $\le$0.01		&$\le$0.1	& 4			& sat		& 0.001		& 	0.0	& 4		\\\cline{2-10}	
			&jain 1 false no overflow					 	& sat 			& $\le$0.01		&$\le$0.1	& 3			& sat 		& 0.007		& 	0.0	& 3		\\\cline{2-10}	
			&jain 2 false no overflow					 	& sat 			& $\le$0.01		&$\le$0.1 	& 3			& sat 		& 0.019		& 	0.0	& 3		\\\cline{2-10}	
			&jain 4 false no overflow	 					& sat 			& $\le$0.01		&$\le$0.1 	& 3			& sat 		& 0.023	& 	0.0	& 3		\\\cline{2-10}		
			&jain 5 false no overflow					 	& sat 			& $\le$0.01		&$\le$0.1	& 3			& sat 		& 0.009		& 	0.0	& 3		\\\cline{2-10}	
			&jain 6 false no overflow					 	& sat 			& $\le$0.01		&$\le$0.1	& 3			& sat 		& 0.024		& 	0.0	& 3		\\\cline{2-10}		
			&jain 7 false no overflow					 	& sat 			&$\le$0.01		&$\le$0.1	& 3			& sat 		& 0.021	& 	0.0	& 3		\\\cline{2-10}	
			&signextension2 false-unreach-call			 	& sat 			& $\le$0.01		&$\le$0.1	& 7			& sat		& 0.002	& 	0.0	& 7		\\\cline{2-10}
			&overflow false unreach call1					& TO		& 600			& 899.0		& /			& TO	& 600			&  366.3		& /		\\\cline{2-10}	
			&overflow false unreach call1 smaller			& sat 			& 11.34			& 113.0		& 507		&sat		& 0.83			& 25.5			& 	502	\\\cline{2-10}		
			&overflow false unreach call1 smaller2			& sat 			& 20.28			& 234.7		& 720		&TO		& 600			& 686.3			& /	\\\cline{2-10}
			&s3 clnt 1 false unreach call true no overflow.BV.c.cil& sat	& 198.93		& 940.5		& 68		& TO		&  600		& 	377		& /		\\\cline{2-10}	
			&s3 clnt 2 false unreach call true no overflow.BV.c.cil& sat	& 0.79			& 57.4		& 4			& sat		&  0.628			& 	33.6		& 4		
			\\
			\hline
			\hline
			\multirow{11}{*}{\rotatebox[origin=c]{90}{no error}}
			&implicitunsignedconversion true unreach call 	& TO 		& 600	& 	300.4	& 13089		& unsat 	& 0.002		& 	0.0 		& 4		\\\cline{2-10}		
			&jain 1 true-unreach-call true-no-overflow	 	& TO 		& 600	& 	1458.1	& 5595		& unsat	& 0.008		& 	0.0 		& 4		\\\cline{2-10}
			&jain 2 true-unreach-call true-no-overflow 		& TO 		& 600	& 	2386.7	& 4230		& unsat	& 0.030 		& 	0.0		& 6		\\\cline{2-10}	
			&jain 4 true-unreach-call true-no-overflow 		& TO 		& 600	&  2795.9	& 3273		& unsat 		& 0.047		& 	0.0	& 6		\\\cline{2-10}	
			&jain 5 true-unreach-call true-no-overflow	 	& TO 		& 600	& 	955.2	& 3083		& unsat	& 0.046		& 	0.0		& 7		\\\cline{2-10}
			&jain 6 true-unreach-call true-no-overflow	 	& TO 		& 600	& 	2447.0	& 2827		& unsat	& 0.052		& 	0.0 		& 6		\\\cline{2-10}
			&jain 7 true-unreach-call true-no-overflow	 	& TO 		& 600	& 	1760.7	& 2731		& unsat	& 0.531		& 	0.0 		& 6		\\\cline{2-10}	
			&signextension2 true-unreach-call			 	& TO 		& 600	& 	504.3	& 11995		& unsat	& 0.002		&   0.0		& 6		\\\cline{2-10}
			&gcd 4 true unreach call true no overflow 		& TO 		& 600	& 	2045	& 2249		& unsat 		& 6.853	& 	25.7 		& 11		\\\cline{2-10}
			&s3 srvr 1 true alt true-no-overflow.BV.c.cil	& TO		& 600 	&	1182.5	& 93		& TO	& 600 		& 	446.5		& /		\\\cline{1-10}				
	\end{tabular}}
\end{table}

\begin{figure}
	\centering
	\includegraphics[width=11cm]{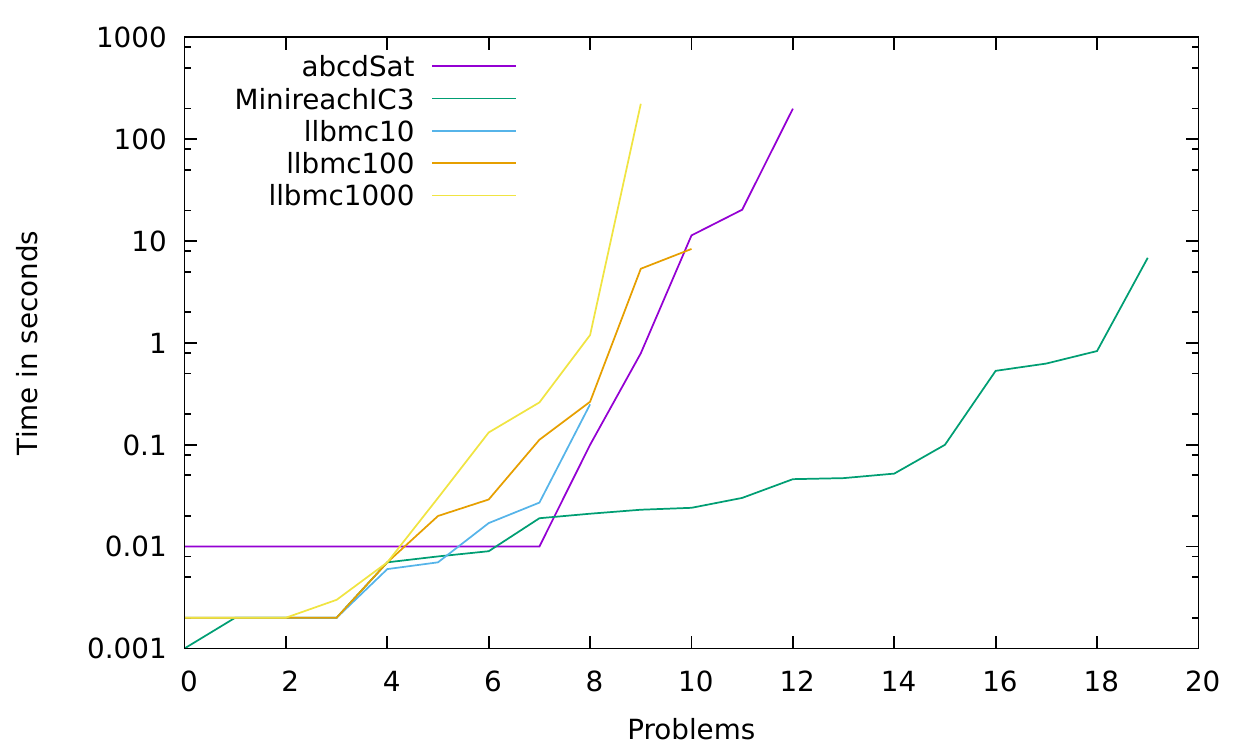}
	\caption{Comparison of LLUMC and LLBMC. The x-axis represents the number of problems the solvers were able to solve and the y-axis the time they needed.}
	\label{fig:fig:compLLUMCLLBMCSecond2}
\end{figure}

MinireachIC3 in comparison is not only able to prove the existence of errors but is also able to prove their non-existence. For erroneous programs the time difference between IncPlan and IC3 is negligible for smaller benchmarks. For some of the larger benchmarks the algorithm produces a timeout.  
In general it is harder to prove the absence of errors than to prove their existence. To prove the existence of an error, the solver only needs to find a valid transition path to the error label, while needing to exclude all possible transition paths to the error label for proving the absence of an error. This complexity is displayed in Table \ref{tab:IncPlan}. The \emph{"jain false"} and \emph{"jain true"} benchmarks only differ in a slightly changed assert statement but to prove the absence of an error always takes more time than to prove its existence. In the case of \emph{"jain 7"} even 25 times longer.

After evaluating the feasibility of our approach Figure \ref{fig:fig:compLLUMCLLBMCSecond2} shows the comparison between the LLUMC-approach with the state-of-the-art bounded model checker LLBMC. When comparing an unbounded model check\-er like LLUMC with a bounded model check\-er, we have to determine a bound until which the bounded model checker unrolls the program. When setting the bound too small, LLBMC runs very fast but has a high chance of producing incorrect results but if we set the bound to high, LLBMC needs a long time to encode and solve the formula. We tested LLBMC with the bounds of 10, 100 and 1000 and compared it with our results generated by IncPlan and MinireachIC3.

Looking at Figure \ref{fig:fig:compLLUMCLLBMCSecond2}, we can recognize the time difference depending on the defined bound. Setting the bound to 10 leads to a really fast solving process but it can solve fewer problems compared to the bound of 100. Setting the bound to 1000 results in timeouts for more complex benchmarks and thus regresses the number of solved problems. After some overhead for smaller problems, solving the benchmarks with IncPlan and abcdSat leads to good results but due to its restriction of only finding errors and not disproving them, it can not solve as many benchmarks as MinireachIC3. The IC3 algorithm can solve 20 out of 24 benchmarks and has a performance advantage compared to all other approaches.

The experimental evaluation illustrates the correctness of our approach for a wide variety of problems. Furthermore, it indicates that the time needed for most problems is reasonable. For model checking in general, the scalability for large programs is always a challenge.

\section{Conclusion and Future Work}
\label{ch:Conclusion}
We introduced a novel unbounded model checking approach to find errors in software or prove their non-existence by using the DimSpec format.
We have developed a new encoding from C-code to a CNF formula in the DimSpec format. Using the intermediate language LLVM, we are able to transform the existence of an error in C-code into four SMT formulas representing the problem of finding a transition path from the initial state of the program to a defined error label. By means of an AIG-supported bit-blasting algorithm, the four SMT formulas are then transformed and added into one CNF in DimSpec format. The encoding has been implemented in the tool LLUMC and we have evaluated this encoding using both the incremental SAT-solving algorithm implemented in the tool IncPlan and the invariant checking algorithm implemented in MinireachIC3. Based on benchmarks from the SV-Comp, the evaluation shows that we extended the functionality of current solvers for infinite-loops while providing correct results and are also comparable to the state-of-the-art solvers regarding solving-time.

Transforming C-code and the existence of errors into CNF formulas in DimSpec format results in a wide range of possibilities to solve the given problem. While we tested incremental SAT-solving and the invariant checking algorithm of IC3, there is also the chance of utilizing advances in parallel SAT-solving for our approach. IncPlan can be run with parallel SAT-solvers as back-end tools, and IC3 was designed to fit both sequential and parallel SAT-solving.

In addition to parallel solving, the performance of the LLUMC approach can also be improved by enlarging the incremental steps of the solver. A first evaluation shows that merging basic blocks in LLVM leads to performance improvements, indicating that a large block encoding could be advantageous.   
Furthermore, the functionality of the  approach can be extended. As a next step, an implementation of other theories like the theory of arrays would make LLUMC usable on a greater range of programs. 

\bibliographystyle{splncs03}
\bibliography{main}

\section*{Appendix}
\label{chap:appendix}
\subsection{Running Example}
\label{sec:LLUMC: Unbounded SBMC:Overview}
To illustrate the transformation from C-code into a LLVM-module and later on into a CNF formula, we demonstrate the encoding  on an example. The example was taken out of the benchmark verification tasks of the competition on software verification (SV-Comp). It can be found under the category \emph{bitvector-loops}.
Example \ref{ex:c-code} iterates through a while-loop until $x$ is smaller then 10. In every loop the value 2 is added to the even number $x$. At the first glance, the loop will never terminate but after a high number of iterations an overflow occurs and the value $x$ becomes smaller then 10, while still being an even number. The maximal value of an unsigned integer ($max\_uint$) is the uneven number $4294967295$. After a high number of loop iterations the value $x$ would be $max\_uint-1$. The addition with $2 \mod (max\_uint+1)$ would then result in $x= max\_uint -1 + 2 - (max\_uint +1) = 0$ and thus the assert condition $(x\%2)$ will fail, because $x$ is still an even number.
This example shows the limitations of bounded model checkers, because they would only unroll the loop to a specific bound that often is not high enough to find errors like these.
\lstset{,
	language=C,
	aboveskip=3mm,
	belowskip=3mm,
	showstringspaces=false,
	columns=flexible,
	basicstyle={\small\ttfamily},
	numbers=left,
	numberstyle=\tiny\color{gray},
	keywordstyle=\textbf,
	commentstyle=\color{gray},
	stringstyle=\color{mauve},
	breaklines=true,
	breakatwhitespace=true,
	tabsize=3
}
\begin{example}[C-Code]\hfill
	\label{ex:c-code}
	\begin{lstlisting}
	int main(void) {
	unsigned int x = 4294967295-101; 
	while (x >= 10) {
	x += 2;
	}		
	__VERIFIER_assert(x % 2);
	}
	\end{lstlisting}
\end{example}
The LLUMC-approach takes this c-Code file and transforms it into an LLVM-module. 

\begin{example}[LLVM-module before optimizations]
	We regard only the main function of the LLVM-module. This function consists of four basic blocks. Variables are marked by an $\%$, representing register variables. The first basic block assigns the constant  4294967194 to the variable $\%x$. It first allocates the needed space before storing the value into the register variable. The second basic block represents the if-condition of Example \ref{ex:c-code}. It loads the value of $x$ into a register variable and the instruction \emph{icmp} checks whether it is greater or equal (uge) then 10. Depending on the output, the branching instruction (\emph{br}) jumps to the third or fourth basic block. The third basic block, representing the body of the if-statement, adds the constant 2 to $x$. The fourth basic block checks the assert condition (x\%2) by first extracting the remainder of the unsigned division of $x$ with 2 (\emph{urem}) and then calling the method $\_\_VERIFIER\_assert$ with the result as a parameter.
	\label{ex:llvm-module}
	\begin{lstlisting}[escapechar=\§]
	define i32 @main() #0 {
	%1 = alloca i32, align 4		
	%x = alloca i32, align 4			§\ \,\framebox[6cm]{unsigned x = 4294967295 - 101;}§
	store i32 0, i32* %1
	store i32 4294967194, i32* %x, align 4
	br label %2
	
	; <label>:2                                       ; preds = %5, %0
	%3 = load i32, i32* %x, align 4
	%4 = icmp uge i32 %3, 10 			§\ \framebox[6cm]{ while(x >= 10)}§
	br i1 %4, label %5, label %8
	
	; <label>:5                                       ; preds = %2
	%6 = load i32, i32* %x, align 4
	%7 = add i32 %6, 2					§\ \framebox[6cm]{ x += 2;}§		
	store i32 %7, i32* %x, align 4
	br label %2
	
	; <label>:8                                       ; preds = %2
	%9 = load i32, i32* %x, align 4
	%10 = urem i32 %9, 2					§\,\,\,\framebox[6cm]{VERIFIER assert(x \% 2);}§			
	call void @__VERIFIER_assert(i32 %10)	
	%11 = load i32, i32* %1
	ret i32 %11
	}
	\end{lstlisting}
\end{example}

In theory one could work on this LLVM-module, but it is more efficient and easier to first run some predefined optimization passes from LLVM and LLBMC. In the first step, we remove \textbf{undefined values} in LLVM. Furthermore, the optimization \textbf{mem2reg} promotes memory references to be register references. The pass called \textbf{inline} tries to inline all functions bottom-up into the main function. Afterwards the two passes \textbf{instnamer} and \textbf{simplifycfg} simplify the program. 
After running these optimizations on our Example \ref{ex:llvm-module} we get the following LLVM-module function as input for the LLUMC-approach. 

\begin{example}[Optimized LLVM-module]
	\label{ex:final llvm module}
	We can see the results of running the LLVM-passes when comparing the resulting main function with the earlier Example \ref{ex:llvm-module}. The result of running the \emph{instname} pass is obvious when looking at the naming of basic blocks and variables. The \emph{mem2reg} pass replaced all allocate, store and load instructions with the \emph{phi} instruction. Hence, the value of $\%x.0$ is set either to  4294967194 when coming from the entry block or to the earlier calculated $\%x +2$. The \emph{inlining} pass inlined the function $\_\_VERIFIER\_assert$ and checks in line 15 whether the assert condition was true(1) or false(0). 
	\begin{lstlisting}[escapechar=\§]
	define i32 @main() #0 
	entry:			
	%tmp = icmp uge i32 4294967194, 10 §\framebox[5.8cm]{unsigned x = 4294967295 - 101;}§
	br i1 %tmp, label %bb1, label %return
	
	bb1:                                              ; preds = %entry, %bb1
	%x.0 = phi i32 [ %tmp2, %bb1 ], [ 4294967194, %entry ]
	%tmp2 = add i32 %x.0, 2			 		 § \ \framebox[5.8cm]{while(x >= 10)\{ x+=2 \}}§
	%tmp3 = icmp uge i32 %tmp2, 10
	br i1 %tmp3, label %bb1, label %return
	
	return:                                           ; preds = %bb1, %entry
	%x.1 = phi i32 [ 4294967194, %entry ], [ %tmp2, %bb1 ]
	%tmp4 = urem i32 %x.1, 2			§\quad\quad\framebox[6cm]{VERIFIER assert(x \% 2);}§
	%tmp.i = icmp ne i32 %tmp4, 0
	br i1 %tmp.i, label %__VERIFIER_assert.exit, label %bb1.i
	
	bb1.i:                                            ; preds = %return
	call void bitcast (void (...)* @__VERIFIER_error to void ()*)() #3		
	unreachable							 	 §\quad\  \framebox[5.8cm]{assert is false}§
	
	__VERIFIER_assert.exit:                           ; preds = %return
	ret i32 0								 §\quad\  \framebox[5.8cm]{assert is true}§
	
	\end{lstlisting}
\end{example}

\subsection{Encoding of the example as described in the paper}
The state space of this example consists of the two variables ${curr}$ and ${pred}$ with a bit-length of four. Furthermore, the variable ${tmp2}$ with a bit-length of 32 is added to the state space, because it occurs in the basic block ${bb1}$ and also in ${return}$. The SMT function $\bvmod$ represents the modulo calculation and the function ${enc(bb)}$ assigns values to the basic block as following: \\

($entry\rightarrow 1,\quad bb1.lr.ph\rightarrow2, \quad bb1\rightarrow3, \quad bb.return\_crit\_edge\rightarrow4,\quad return\rightarrow5,\quad bb.1\rightarrow6,\quad \_\_VERIFIER\_assert\_exist\rightarrow7,\quad ok\rightarrow8,\quad error\rightarrow9$).\\

The encoding algorithm iterates over all basic blocks of the LLVM-module and encodes them as described in the paper. The encoding of the example leads to the following formulas, which are then transformed to CNF-formulas by an AIG-based approach. \\

\noindent \textbf{Initial Formula:}
\begin{alignat*}{2}
curr 	&= 1\quad \land  \\
pred	&= 1\,.
\end{alignat*}

\noindent \textbf{Goal Formula:}
\begin{alignat*}{2}
curr 	&= 9 \,.
\end{alignat*}

\noindent \textbf{Universal Formula:}
\begin{alignat*}{2}
curr	&\leq 9 \quad \land \\
pred	&\leq 9 \,.
\end{alignat*}

\noindent \textbf{Transition Formula:}

\begin{alignat*}{2}
&(curr = 1 \land pred = 1 	\Rightarrow curr'=9 \land pred'=9 \land tmp2'=tmp2)																		\\
&\land \quad (curr=2 					\Rightarrow curr'=3 \land pred'=2 \land tmp2'=tmp2)																		\\
&\land \quad (curr=3 \land 10 \leq (2+{\ite}(pred=2, 10, tmp2))	\\
&\qquad\qquad\Rightarrow curr'=3 \land pred'=3 \land tmp2'=(2+{\ite}(pred=2, 10, tmp2)))		\\
&\land \quad (curr=3 \land 10 > (2+{\ite}(pred=2, 10, tmp2))		\\
&\qquad\qquad \Rightarrow curr'=4 \land pred'=3 \land tmp2'=(2+{\ite}(pred=2, 10, tmp2)))\\
&\land \quad (curr = 4 					\Rightarrow curr'=5 \land pred'=4 \land tmp2'=tmp2	)																	\\
&\land \quad (curr=5 \land 0 \neq {\bvmod}({\ite}(pred=4, tmp2, 10), 2)	\\
&\qquad\qquad\Rightarrow curr'=7 \land pred'=5 \land tmp2'=tmp2)							\\
&\land \quad (curr=5 \land 0 =  {\bvmod}({\ite}(pred=4, tmp2, 10), 2)	\\
&\qquad\qquad\Rightarrow curr'=6 \land pred'=5 \land tmp2'=tmp2)							\\
&\land \quad (curr=6 					\Rightarrow curr'=9 \land pred'=6 \land tmp2'=tmp2	)																	\\
&\land \quad (curr=7 					\Rightarrow curr'=8 \land pred'=7 \land tmp2'=tmp2	)																	\\
&\land \quad (curr=8 					\Rightarrow curr'=8 \land pred'=8 \land tmp2'=tmp2	)																\\
&\land \quad (curr=9 					\Rightarrow curr'=9 \land pred'=7 \land tmp2'=tmp2 	)																	
\end{alignat*}

\subsection{Details from the Experimental Evaluation}
Details about the benchmarks used for the experimental evaluation are given in table format. Furthermore, detailed evaluation results are displayed. 
\begin{table}[h]
	\caption{Selected Benchmarks, 
		where the column LoC stands for the lines of code, \#Variables presents the number of SMT variables, \#Bits displays the number of SAT variables and \#Clauses the number of clauses generated by our approach.}
	\bigskip
	\label{tab:bData}
	\centering
	\Rotatebox{90}{%
		\begin{tabular}{|l|l||c|c|c|c|c|c|}
			\cline{2-8}
			\multicolumn{1}{c|}{}& {\textbf{Benchmark}} 				& \textbf{Desired Result} & \textbf{LoC} & \textbf{\#Basic Blocks}& \textbf{\#Variables}& \textbf{\#Bits} & \textbf{\#Clauses} \\\cline{2-8} \cline{1-8}	
			\multirow{13}{*}{\rotatebox[origin=c]{90}{error}}
			&diamond false unreach call2 					& sat			& 48			& 	37	& 34	&1036	&35788\\\cline{2-8}	
			&implicitunsignedconversion false unreach call	& sat 		& 14			&   4	& 2		&6		&151	\\\cline{2-8}	
			&jain 1 false no overflow					 	& sat 		& 25			&	5	& 3		&38		&2179	\\\cline{2-8}		
			&jain 2 false no overflow					 	& sat 		& 26		 	& 	5	& 4		&70		&5125	\\\cline{2-8}	
			&jain 4 false no overflow	 					& sat 		& 27			&	5 	& 5		&102	&8022	\\\cline{2-8}		
			&jain 5 false no overflow					 	& sat 		& 27			&	5	& 4		&70		&3415	\\\cline{2-8}	
			&jain 6 false no overflow					 	& sat 		& 27			&	5	& 5		&102	&7999	\\\cline{2-8}	
			&jain 7 false no overflow					 	& sat 		& 27 			& 	5	& 5		&102	&7171	\\\cline{2-8}	
			&signextension2 false-unreach-call			 	& sat 		& 19			& 	7	& 2		&8		&263	\\\cline{2-8}
			&overflow false unreach call1					& sat 		& 18			& 	7	& 3		&40		&1652	\\\cline{2-8}	
			&overflow false unreach call1 smaller			& sat 		& 18			& 	7	& 3		&40		&1652	\\\cline{2-8}	
			&overflow false unreach call1 smaller2			& sat 		& 18			& 	7	& 3		&40		&1637	\\\cline{2-8}	
			&s3 clnt 1 false unreach call true-no-overflow.BV.c.cil& sat& 646			& 	227	& 93	&2928	&116777	\\\cline{2-8}
			&s3 clnt 2 false unreach call true-no-overflow.BV.c.cil& sat& 624			& 	227	& 95	&2992	&94826	\\\cline{1-8}
			\multirow{11}{*}{\rotatebox[origin=c]{90}{no error}}
			&implicitunsignedconversion true unreach call 	& unsat 	& 14			& 	4 	& 2		&6		&151	\\\cline{2-8}
			&jain 1 true-unreach-call true-no-overflow	 	& unsat 	& 25			& 	5	& 3		&38		&2128	\\\cline{2-8}
			&jain 2 true-unreach-call true-no-overflow 		& unsat 	& 26			& 	5	& 4		&70		&4999	\\\cline{2-8}
			&jain 4 true-unreach-call true-no-overflow 		& unsat 	& 27			&  	5	& 5		&102	&7738	\\\cline{2-8}
			&jain 5 true-unreach-call true-no-overflow	 	& unsat  	& 27			& 	5	& 4		&70		&3370	\\\cline{2-8}
			&jain 6 true-unreach-call true-no-overflow	 	& unsat 	& 27			& 	5	& 5		&102	&7672\\\cline{2-8}	
			&jain 7 true-unreach-call true-no-overflow	 	& unsat 	& 27			& 	5	& 5		&102	&6148 \\\cline{2-8}			
			&signextension2 true-unreach-call			 	& unsat 	& 19			& 	7	& 2		&8		&263	\\\cline{2-8}
			&gcd 4 true unreach call true no overflow 		& unsat 	& 46			& 	12	& 8		&200	&21862	\\\cline{2-8}
			&s3 srvr 1 true alt true-no-overflow.BV.c.cil	& unsat 	& 696			& 	256	& 99	&3122	&102932	\\\cline{1-8}
	\end{tabular}}
\end{table}

\begin{table}[h]
	\caption{Runtime data for Benchmarks from the SV-Comp run with IncPlan, where the column Iterations stands for the number of incremental solving steps performed by the SAT-solver and the result SEG shows the occurrence of an segmentation fault. }
	\bigskip
	\label{tab:IncPlan}
	\centering
	\resizebox{8.5cm}{!}{
		\Rotatebox{90}{%
			\begin{tabular}{|l|l|c|c|c|c||c|c|c|c|}
				\cline{3-10}
				\multicolumn{2}{c|}{} & \multicolumn{4}{c||}{{abcdSAT}} & \multicolumn{4}{c|}{{PicoSat}}\\
				\cline{2-10}
				\multicolumn{1}{c|}{} &	\textbf{Benchmark} 	& \textbf{Result} 	 	& \textbf{Time/s} & \textbf{Memory/MB}& \textbf{Iterations} & \textbf{Result} 	& \textbf{Time/s} 	& 	\textbf{Memory/MB} 	& \textbf{Iterations}\\ \cline{2-10} \cline{1-10}
				\multirow{13}{*}{\rotatebox[origin=c]{90}{error}}
				&diamond false unreach call2 					& sat 			& 3.95& 	130		& 27				& sat 		& 4.02			& 	69.5 		& 27	\\\cline{2-10}	
				&implicitunsignedconversion false unreach call	& sat 			& $\le$0.01		&$\le$0.1	& 4			& sat		& $\le$0.01		& 	$\le$0.1	& 4		\\\cline{2-10}	
				&jain 1 false no overflow					 	& sat 			& $\le$0.01		&$\le$0.1	& 3			& sat 		& $\le$0.01		& 	$\le$0.1	& 3		\\\cline{2-10}	
				&jain 2 false no overflow					 	& sat 			& $\le$0.01		&$\le$0.1 	& 3			& sat 		& $\le$0.01		& 	$\le$0.1	& 3		\\\cline{2-10}	
				&jain 4 false no overflow	 					& sat 			& $\le$0.01		&$\le$0.1 	& 3			& sat 		& $\le$0.01		& 	$\le$0.1	& 3		\\\cline{2-10}		
				&jain 5 false no overflow					 	& sat 			& $\le$0.01		&$\le$0.1	& 3			& sat 		& $\le$0.01		& 	$\le$0.1	& 3		\\\cline{2-10}	
				&jain 6 false no overflow					 	& sat 			& $\le$0.01		&$\le$0.1	& 3			& sat 		& $\le$0.01		& 	$\le$0.1	& 3		\\\cline{2-10}		
				&jain 7 false no overflow					 	& sat 			&$\le$0.01		&$\le$0.1	& 3			& sat 		& $\le$0.01		& 	$\le$0.1	& 3		\\\cline{2-10}	
				&signextension2 false-unreach-call			 	& sat 			& $\le$0.01		&$\le$0.1	& 7			& sat		& $\le$0.01		& 	$\le$0.1	& 7		\\\cline{2-10}
				&overflow false unreach call1					& timeout		& 600			& 899.0		& /			& timeout	& 600			&  483.3		& /		\\\cline{2-10}	
				&overflow false unreach call1 smaller			& sat 			& 11.34			& 113.0		& 507		&sat		& 10.08			& 69.1			& 	507	\\\cline{2-10}		
				&overflow false unreach call1 smaller2			& sat 			& 20.28			& 234.7		& 720		&sat		& 24.35			& 194.3			& 720	\\\cline{2-10}
				&s3 clnt 1 false unreach call true no overflow.BV.c.cil& sat	& 198.93		& 940.5		& 68		& sat		&  421.57		& 	530.6		& 68		\\\cline{2-10}	
				&s3 clnt 2 false unreach call true no overflow.BV.c.cil& sat	& 0.79			& 57.4		& 4			& sat		&  0.8			& 	42.4		& 4		\\\cline{1-10}
				\multirow{11}{*}{\rotatebox[origin=c]{90}{no error}}
				&implicitunsignedconversion true unreach call 	& timeout 		& 600	& 	300.4	& 13089		& timeout 	& 600		& 	281.7 		& 17076		\\\cline{2-10}		
				&jain 1 true-unreach-call true-no-overflow	 	& timeout 		& 600	& 	1458.1	& 5595		& timeout	& 600		& 	704.2 		& 5477		\\\cline{2-10}
				&jain 2 true-unreach-call true-no-overflow 		& timeout 		& 600	& 	2386.7	& 4230		& timeout	& 600 		& 	1072.7 		& 3731		\\\cline{2-10}	
				&jain 4 true-unreach-call true-no-overflow 		& timeout 		& 600	&  2795.9	& 3273		& timeout 		& 600		& 	1246.3		& 2832		\\\cline{2-10}	
				&jain 5 true-unreach-call true-no-overflow	 	& timeout 		& 600	& 	955.2	& 3083		& timeout	& 600		& 	497.4 		& 2617		\\\cline{2-10}
				&jain 6 true-unreach-call true-no-overflow	 	& timeout 		& 600	& 	2447.0	& 2827		& timeout	& 600		& 	1252.6 		& 2844		\\\cline{2-10}
				&jain 7 true-unreach-call true-no-overflow	 	& timeout 		& 600	& 	1760.7	& 2731		& timeout	& 600		& 	784.4 		& 1406		\\\cline{2-10}	
				&signextension2 true-unreach-call			 	& timeout 		& 600	& 	504.3	& 11995		& timeout	& 600		&   496.0		& 16901		\\\cline{2-10}
				&gcd 4 true unreach call true no overflow 		& timeout 		& 600	& 	2045	& 2249		& SEG 		& 104.03	& 	3750.8 		& 2103		\\\cline{2-10}
				&s3 srvr 1 true alt true-no-overflow.BV.c.cil	& timeout		& 600 	&	1182.5	& 93		& timeout	& 600 		& 	495.7		& 78		\\\cline{1-10}				
	\end{tabular}}}
\end{table}

\begin{table}[h]
	\caption{Runtime data for Benchmarks from the SV-Comp run with MinireachIC3, where the column Phases represents the number of major steps performed by the solver.}
	\bigskip
	\label{tab:IC3}
	\centering
	\resizebox{12cm}{!}{
		\begin{tabular}{|l|l||c|c|c|c|}
			\cline{2-6}
			\multicolumn{1}{c|}{} & \textbf{Benchmark} 		& \textbf{Result} 	 	& \textbf{Time/s} & \textbf{Memory/MB}& \textbf{Phases}\\\cline{1-6} \cline{2-6}
			\multirow{13}{*}{\rotatebox[origin=c]{90}{error}}
			&diamond false unreach call2 					& timeout 		& 600			& 	241.2	& /	\\\cline{2-6}
			&implicitunsignedconversion false unreach call	& sat 			& 0.001			&   0.0		& 3		\\\cline{2-6}	
			&jain 1 false no overflow					 	& sat 			& 0.007			&	0.0		& 2		\\\cline{2-6}	
			&jain 2 false no overflow					 	& sat 			& 0.019		 	& 	0.0		& 2		\\\cline{2-6}	
			&jain 4 false no overflow	 					& sat 			& 0.023			&	0.0 	& 2		\\\cline{2-6}		
			&jain 5 false no overflow					 	& sat 			& 0.009			&	0.0		& 2		\\\cline{2-6}	
			&jain 6 false no overflow					 	& sat 			& 0.024			&	0.0		& 2		\\\cline{2-6}	
			&jain 7 false no overflow					 	& sat 			& 0.021			& 	0.0		& 2		\\\cline{2-6}		
			&signextension2 false-unreach-call			 	& sat 			& 0.002			& 	0.0		& 5		\\\cline{2-6}	
			&overflow false unreach call1 					& timeout 		& 600			& 	366.3	& /		\\\cline{2-6}
			&overflow false unreach call1 smaller			& sat 			& 0.83			& 	25.5	& 502	\\\cline{2-6}	
			&overflow false unreach call1 smaller2			& timeout 		& 600			& 	686.3	& /		\\\cline{2-6}	
			&s3 clnt 1 false unreach call true no overflow.BV.c.cil& timeout& 600			& 	377		& /		\\\cline{2-6}		
			&s3 clnt 2 false unreach call true no overflow.BV.c.cil& sat	& 0.628			& 	33.6	&3		\\\cline{1-6}
			\multirow{11}{*}{\rotatebox[origin=c]{90}{no error}}
			&implicitunsignedconversion true unreach call 	& unsat 		& 0.002			& 	0.0 	& 3		\\\cline{2-6}
			&jain 1 true-unreach-call true-no-overflow	 	& unsat 		& 0.008			& 	0.0		& 5		\\\cline{2-6}
			&jain 2 true-unreach-call true-no-overflow 		& unsat 		& 0.030			& 	0.0		& 5		\\\cline{2-6}
			&jain 4 true-unreach-call true-no-overflow 		& unsat 		& 0.047			&  	0.0		& 5		\\\cline{2-6}	
			&jain 5 true-unreach-call true-no-overflow	 	& unsat	 		& 0.046			& 	0.0		& 6		\\\cline{2-6}	
			&jain 6 true-unreach-call true-no-overflow	 	& unsat 		& 0.052			& 	0.0		& 5		\\\cline{2-6}
			&jain 7 true-unreach-call true-no-overflow	 	& unsat 		& 0.531			& 	0.0		& 5		\\\cline{2-6}	
			&signextension2 true-unreach-call			 	& unsat 		& 0.002			& 	0.0		& 5		\\\cline{2-6}
			&gcd 4 true unreach call true no overflow 		& unsat			& 6.853			& 	25.7	& 10	\\\cline{2-6}
			&s3 srvr 1 true alt true-no-overflow.BV.c.cil	& timeout		& 600 			&   446.5	&/		\\\cline{1-6}							
	\end{tabular}}
\end{table}

\begin{table}[!h]
	\caption{Comparison between LLUMC and LLBMC.}
	\bigskip
	\label{tab:LLBMC}
	\centering
	\resizebox{8.5cm}{!}{
		\Rotatebox{90}{%
			
			\begin{tabular}{|l|l|c|c|c|c|c|c||c|c|c|c|}
				\cline{3-12}
				\multicolumn{2}{c|}{} & \multicolumn{6}{c||}{{LLBMC}} & \multicolumn{4}{c|}{{LLUMC}}\\
				\cline{2-12}
				\multicolumn{1}{c|}{} 	&\textbf{Benchmark} 				& \textbf{Bound 10}	& \textbf{Time/s}& \textbf{Bound 100}& \textbf{Time/s}& \textbf{Bound 1000}& \textbf{Time/s} & \textbf{abcdSat}& \textbf{Time/s} & \textbf{MinireachIC3}& \textbf{Time/s}\\\cline{2-12} \cline{1-12}	
				\multirow{13}{*}{\rotatebox[origin=c]{90}{error}}
				&diamond false unreach call2 					& sat	& 0.017	& sat	& 0.112& sat	& 1.194 & sat	& 0.1	& sat	& 0.1	\\\cline{2-12}	
				&implicitunsignedconversion false unreach call	& sat	& 0.002	& sat	& 0.002& sat	& 0.002	& sat	& 0.01 & sat	& 0.001		\\\cline{2-12}	
				&jain 1 false no overflow					 	& i.b.	& 0.008	& i.b.	& 0.031& i.b.	& 0.237 & sat	& 0.01 & sat	& 0.007	\\\cline{2-12}		
				&jain 2 false no overflow					 	& i.b.	& 0.009	& i.b.	& 0.038& i.b.	& 0.316	& sat	& 0.01 & sat	& 0.019			\\\cline{2-12}
				&jain 4 false no overflow	 					& i.b.	& 0.010	& i.b.	& 0.044& i.b.	& 0.381	& sat & 0.01 & sat	& 0.023	\\\cline{2-12}		
				&jain 5 false no overflow					 	& i.b.	& 0.005	& i.b.	& 0.019& i.b.	& 0.134	& sat	& 0.01 & sat	& 0.009		\\\cline{2-12}	
				&jain 6 false no overflow					 	& i.b.	& 0.012	& i.b.	& 0.051& i.b.	& 0.429	& sat	& 0.01 & sat	& 0.024	\\\cline{2-12}		
				&jain 7 false no overflow					 	& i.b.	& 0.008	& i.b.	& 0.054& i.b.	& 0.449	& sat	& 0.01 & sat	& 0.021	\\\cline{2-12}		
				&signextension2 false-unreach-call			 	& sat	& 0.002	& sat	& 0.002& sat	& 0.003 & sat	& 0.01 & sat	& 0.002		\\\cline{2-12}
				&overflow false unreach call1 			& sat	& 0.007	& sat	& 0.007& sat	& 0.007 & timeout& 600  & timeout	& 600		\\\cline{2-12}	
				&overflow false unreach call1 smaller			& sat	& 0.027	& sat	& 0.029& sat	& 0.030 & sat	& 11.34	& sat	& 0.83	\\\cline{2-12}	
				&overflow false unreach call1 smaller2			& sat	& 0.249 & sat	& 0.264	& sat	& 0.260 & sat	& 20.28	& timeout	& 600	\\\cline{2-12}	
				&s3 clnt 1 false unreach call true no overflow.BV.c.cil& i.b.	& 0.338	& sat	& 8.347	& timeout	& 600 & sat	& 198.93 & timeout	& 600	\\\cline{2-12}
				&s3 clnt 2 false unreach call true no overflow.BV.c.cil& i.b.	& 0.295	& sat	& 5.336	& sat	& 222.414 & sat	& 0.79 & sat	& 0.628	\\\cline{1-12}
				\multirow{11}{*}{\rotatebox[origin=c]{90}{no error}}									
				
				&implicitunsignedconversion true unreach call 	& no error& 0.002& no error& 0.002& no error	& 0.002		& timeout& 600& unsat	& 0.002		\\\cline{2-12}	
				&jain 1 true-unreach-call true-no-overflow	 	& i.b. & 0.008	& i.b.	& 0.031& i.b.	& 0.239	   	& timeout& 600& unsat	& 0.008		\\\cline{2-12}	
				&jain 2 true-unreach-call true-no-overflow 		& i.b. & 0.009	& i.b.	& 0.039& i.b.	& 0.316		& timeout& 600& unsat	& 0.030	   \\\cline{2-12}
				&jain 4 true-unreach-call true-no-overflow 		& i.b. & 0.011	& i.b.	& 0.045& i.b.	& 0.370		& timeout& 600& unsat	& 0.047		\\\cline{2-12}	
				&jain 5 true-unreach-call true-no-overflow	 	& i.b. & 0.005	& i.b.	& 0.020& i.b.	& 0.128		& timeout& 600& unsat	& 0.046	\\\cline{2-12}
				&jain 6 true-unreach-call true-no-overflow	 	& i.b. & 0.010	& i.b.	& 0.051& i.b.	& 0.407		& timeout& 600& unsat	& 0.052	\\\cline{2-12}
				&jain 7 true-unreach-call true-no-overflow	 	& i.b. & 0.013	& i.b.	& 0.054& i.b.	& 0.451		& timeout& 600& unsat	& 0.531		\\\cline{2-12}	
				&signextension2 true-unreach-call			 	& no error & 0.002	& no error	& 0.002& no error	& 0.002	& timeout& 600& unsat	& 0.002	\\\cline{2-12}	
				&gcd 4 true unreach call true no overflow 		& no error & 0.006	& no error	& 0.020& no error	& 0.132	& timeout& 600& unsat	& 6.853	\\\cline{2-12}	
				&s3 srvr 1 true alt true-no-overflow.BV.c.cil	& i.b. & 1.802	& i.b.	& 266.09 & i.b.	& 600		& timeout& 600& timeout	& 600	\\\cline{1-12}	
	\end{tabular}}}
\end{table}

\end{document}